\documentclass[fleqn,10pt]{wlscirep}
\usepackage{amsmath}
\usepackage[utf8]{inputenc}
\usepackage[T1]{fontenc}
\usepackage{color}
\usepackage{bm}
\usepackage{array}
\usepackage{tcolorbox}
\usepackage{makecell}
\usepackage{xcolor,colortbl}
\usepackage{float}
\usepackage{tabu}
\usepackage{ulem}

\newcommand{\hbarhorizline}{\mathchar'26\mkern-7mu h}

\title{Advances in Quantum Imaging}

\definecolor{lightgreen}{RGB}{236, 240, 231} 
\definecolor{darkgreen}{RGB}{221, 229, 213}

\newtcolorbox{mybox}{
    colback=lightgreen,  
    colframe=white,           
}

\author[1,*]{Hugo Defienne}
\author[2,3]{Warwick P. Bowen}
\author[4]{Maria Chekhova}
\author[5]{Gabriela Barreto Lemos}
\author[6]{Dan Oron}
\author[7,8]{Sven Ramelow}
\author[9]{Nicolas Treps}
\author[10]{Daniele Faccio}

\affil[1]{Sorbonne Université, CNRS, Institut des NanoSciences de Paris, INSP, F-75005 Paris, France}
\affil[2]{School of Mathematics and Physics, The University of Queensland, St Lucia, Queensland 4067, Australia}%
\affil[3]{ARC Centre of Excellence in Quantum Biotechnology, St Lucia, Queensland 4067, Australia}
\affil[4]{Max Planck Institute for the Science of Light, Staudtstraße 2, 91058 Erlangen, Germany}
\affil[5]{Instituto de Física, Universidade Federal do Rio de Janeiro, Av. Athos da Silveira Ramos 149, Rio de Janeiro, CP 68528, Brazil}
\affil[6]{Department of Molecular Chemistry and Materials Science,Weizmann Institute of Science, Rehovot 7610001, Israel}
\affil[7]{
Institut für Physik, Humboldt-Universität zu Berlin, Newtonstr. 15, 12489 Berlin, Germany}
\affil[8]{IRIS Adlershof, Humboldt-Universität zu Berlin, Berlin, Germany}
\affil[9]{Laboratoire Kastler Brossel, ENS-Universite PSL, CNRS, Sorbonne Universite, College de France, 24 rue Lhomond, 75005 Paris, France}%
\affil[10]{School of Physics and Astronomy,University of Glasgow, Glasgow G12 8QQ, UK}%

\affil[*]{e-mail: hugo.defienne@insp.upmc.fr}

\begin{abstract}
Modern imaging technologies are widely based on classical principles of light or electromagnetic wave propagation. They can be remarkably sophisticated, with recent successes ranging from single molecule microscopy to imaging far-distant galaxies. However, new imaging technologies based on quantum principles are gradually emerging. They can either surpass classical approaches or provide novel imaging capabilities that would not otherwise be possible. {Here }we provide an overview  {of the most recently developed quantum imaging systems, highlighting the non-classical properties of sources such as bright squeezed light, entangled photons, and single-photon emitters that enable their functionality.} We outline potential upcoming trends and the associated challenges, all driven by a central inquiry, which is to understand whether quantum light can make visible the invisible.
\end{abstract}
\begin{document}

\flushbottom
\maketitle

\thispagestyle{empty}

From microscopy, {delving into} the complex microcosms of biology, to the far reaches of astronomy, where we {marvel} at the cosmos, imaging {serves as our lens for exploring and understanding the} universe. However, despite the remarkable capabilities of modern imaging systems, {they encounter} limitations stemming from technical and fundamental boundaries. A prime example is spatial resolution~\cite{lupo2016ultimate}.
Once a microscope is technically optimized, its ability to resolve the smallest details is constrained by {fundamental limits including diffraction (due to light's wave-like nature) and shot noise (related to light's particle-like behavior)}. Improving resolution thus requires reducing one or both, necessitating an exploration of light's fundamental nature at its core. The quantum limit of information content carried by light can only be computed taking into account these two aspects, and shows that classical criteria are orders of magnitude away from the fundamental limits.
{In biology, this could mean observing and understanding previously invisible microscopic mechanisms, while in astronomy, where sources are inaccessible and photon numbers are limited, it could enable surpassing Rayleigh's criterion to resolve distant stars.} 
These are some of the exciting prospects that quantum imaging, and more broadly, quantum-informed imaging protocols bring to the table — a domain where quantum phenomena and optics combine to reduce the boundaries of what was once deemed impossible. 

Expanding upon quantum metrology and sensing~\cite{giovannetti2011advances}, the notion of utilizing quantum {light properties} to {enhance imaging} emerged {in the last} 40 years. Technological advances in detectors and sources, coupled with conceptual progress {in quantum phenomena like entanglement and non-locality}, have progressively led towards applications.  In the past five years, we have for instance witnessed the first {proof-of-principle} demonstrations of quantum microscopes  utilizing squeezed light~\cite{casacio2021quantum}, single-photon emitters~\cite{tenne2019super} and entangled photons~\cite{kviatkovsky2020microscopy}. {Despite their} limited scope, some of {these devices} enable the observation of {otherwise indiscernible} structures~\cite{casacio2021quantum} . {Beyond bioimaging,} fields such as remote sensing~\cite{lloyd2008enhanced}, material science~\cite{boto2000quantum} and astronomy~\cite{crawford2023towards} {also} benefit from {quantum imaging} research and advancements. 

However, while quantum imaging holds promise, practical implementation {faces challenges}. Many protocols {are hindered} by the very low intensity of non-classical sources and the precariousness of the benefits they yield. {Today' researchers strive to develop robust quantum imaging technologies for real-world use. }
This review {aims} to highlight recent advances in quantum imaging, focus{ing} on {developments poised to practical use, notably in optical microscopy.}
Readers {can refer} to previous reviews in the field~\cite{kolobov2007quantum,genovese2016real,moreau2019imaging,gatti2008quantum,gilaberte2019perspectives,magana2019quantum,shih2007quantum,lubin2022photon,moodley2024advances} {and} the historical overview in Box 1 for additional context. 
 
\begin{mybox}

\paragraph{Box 1 | Brief history of quantum imaging } 
\label{History}
Basic types of nonclassical light {emerged during} the 1960's – 1980's{.} {These encompassed} entangled photons emitted through cascaded transitions in atoms~\cite{kocher1967} or spontaneous parametric down-conversion (SPDC)~\cite{burnham1970observation}, single photons emitted by atoms~\cite{kimble1977}, {and} squeezed light obtained {via} four-wave mixing~\cite{slusher1985} and SPDC~\cite{xiao1987precision}. Soon after that, various ideas for the applications of nonclassical light {arose} - most notably, for imaging. 

One {remarkable aspect} was the noise reduction {achieved by} squeezed light. Sub-shot-noise fluctuations of the photon number enable both low-noise imaging and improvements in spatial resolution~\cite{kolobov1991}. In the first experiment demonstrating enhanced resolution, a split detector registered the position of a squeezed-light beam better than  a coherent beam~\cite{treps2002surpassing}. Later, it was understood that a flux of single photons or any other sub-Poissonian light {could achieve similar results}, {leading to} experiments with single-photon emitters~\cite{schwartz2013superresolution,monticone2014beating}. {This was facilitated by} the discovery of solid-state ‘artificial atoms’: color centers in diamond, single molecules, and quantum dots.

Meanwhile, photon pairs offered another rich resource for imaging. The first idea, ghost imaging~\cite{pittman1995optical}, used position-momentum entanglement between the signal and idler photons of a pair. To image an object in the signal channel, it was sufficient to measure correlations between the signal photons registered by a `bucket’ detector and a spatially-resolving detector in the idler channel, which contained no object (Fig.~B1.a). The same technique was later implemented using sources with classical intensity correlations~\cite{bennink2002,valencia2005}, albeit with a lower image contrast. 

Further, Boto et al.~\cite{boto2000quantum} {highlighted} another striking feature of a photon pair: it behaves as a single quantum object with the frequency equal to the sum of signal and idler frequencies. Therefore, as demonstrated in Ref.~\cite{d2001two}, it shows a diffraction pattern with a reduced wavelength and could be used to create sharper images.

In 1977, Klyshko~\cite{klyshko1977} suggested that beams consisting of pairs of distinguishable photons - twin beams – provide a light source with a given number of photons if detection of one photon is used to ‘herald’ the arrival of the other. Later, such sources were demonstrated experimentally~\cite{hong1986,rarity1987}. This opened a path for obtaining brighter sub-Poissonian beams by detecting one of the twin beams and imposing a condition on its photon number~\cite{laurat2003}. Spatially multimode sub-Poissonian beams, prepared {similarly}~\cite{iskhakov2016}, are {well-suited} for sub-shot-noise imaging.

But is it necessary to prepare a sub-Poissonian beam {beforehand} for imaging? Brambilla et al. proposed~\cite{brambilla2008} and Brida et al. implemented~\cite{brida2010experimental} sub-shot-noise imaging {using} twin beams: the object is placed into one beam, the resulting images are subtracted, and the noise is reduced due to the photon-number correlations (Fig.~B1.b). 

{In 1991,} a very different effect with entangled photons, {termed} induced coherence, was first demonstrated~\cite{wang1991}: if photon pairs are generated by two coherently pumped but spatially separated sources, the signal photons from both sources become coherent whenever their idler photons {share} a common path. {Placing an object} between these sources partly breaks this coherence{, allowing imaging} by registering only signal photons~\cite{lemos2014quantum} (Fig.~B1.c).

\begin{figure}[H]
\centering
\includegraphics[width=1\linewidth]{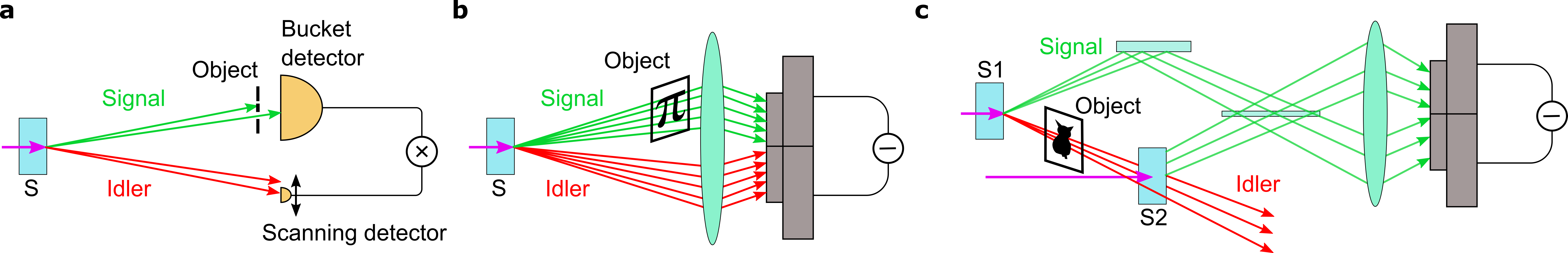}
\caption*{\textbf {Figure B1. Examples of quantum imaging schemes.} {\textbf{a,}} Ghost imaging~\cite{pittman1995optical}. A source S emits pairs of spatially entangled photons. Signal photons pass through an absorptive object and are then collected by a `bucket' detector. The object shape is recovered by scanning a spatially-resolving detector in the idler channel and registering signal-idler correlations. {\textbf{b,}} Sub-shot-noise imaging with twin beams~\cite{brida2010experimental}. A weakly absorbing object is placed in the signal beam and the images of both beams are detected by cameras and subtracted. The image noise is then reduced due to the signal-idler photon-number correlations. {\textbf{c,}} Imaging with undetected photons~\cite{lemos2014quantum}. Coherently pumped sources S1 and S2 generate photon pairs, so that idler photons from S1 pass through S2. Signal photons from both sources are overlapped on a beam splitter and they interfere if their idler partners are indistinguishable. An object placed in the idler channel between the two sources breaks this indistinguishability, {allowing} a camera in the signal channel {to} reveal the shape of the object.} 
\label{Figure1}
\end{figure}

\end{mybox}

\section{Classical and quantum bounds on image resolution}
\label{part0}

When dealing with resolution, or measurement sensitivity, benchmarking is {crucial.} Resolution is bounded by measurement design, detectors sensitivity, light source noise and, ultimately, the quantum nature of light. Quantum metrology provides the framework to compute the ultimate precision limit on the estimation of any parameter encoded in a beam of light. The Cram\'er-Rao bound~\cite{cramer1999mathematical,rao1947minimum} indicates the minimum variance of any unbiased estimator for a given measurement setting. The Quantum Cram\'er-Rao bound (QCRB)~\cite{caves1981quantum,helstrom1969quantum} is the optimization of the Cram\'er Rao bound over all measurements allowed by quantum mechanics, {providing} a limit independent of the measurement strategy. It {is} calculated as the inverse of the Quantum Fisher Information, {requiring understanding} how the input light quantum state varies with the parameter under consideration~\cite{giovannetti2011advances}.

\paragraph{Optimal measurement.} While mathematically involved in the general case, a simple situation arises when the parameter is imprinted via a unitary transformation with a parameter-independent Hamiltonian. {In this situation, }for pure states, one has 
$i\ \hbarhorizline\frac{\partial |\psi\rangle}{\partial\theta} = \hat H|\psi\rangle$ and the quantum Fisher information is $F_Q[|\psi\rangle,\hat H] = 4\Delta^2\hat H_{|\psi\rangle}$. The variance of any estimator is {thus} bounded by the inverse of the variance of the Hamiltonian governing the state evolution under the parameter $\theta$~\cite{braunstein1994statistical}. {For} example, this is why NOON states{,} $\frac{1}{\sqrt{2}}(|N,0\rangle + |0,N\rangle)${,} that are superpositions of $N$-photons Fock states and vacuum states {over two arms of an interferometer}, are optimal for phase measurement~\cite{pezze2018quantum}. This state maximizes the energy variance when the energy observable is the Hamiltonian for phase evolution. 

In practical applications, however, optimizing a measurement cannot be reduced to finding the optimal state as this will often be impractical. First, sensitivity or standard deviation of the estimator is always proportional to some power of the inverse of the number of detected photons. {Increasing photon flux is thus the primary} and most efficient strategy{, with different scaling depending} on the light source:  with coherent states one finds the usual shot noise limit $\propto \frac{1}{\sqrt{N}}$, while using NOON states standard deviation scales as $ \frac{1}{{N}}$. Squeezed states, {though theoretically capable of} $\propto \frac{1}{\sqrt{N}}$ scaling, are mainly used in situations where they multiply the global sensitivity by a fixed factor, equal to the squeezing factor~\cite{caves1981quantum}. This has the advantage {of being} compatible with the large photon number regime, {thereby improving} practical scenarios. Furthermore, the general sensitivity, especially in imaging, {relies} on how the mean classical field {or} image varies {with} the parameter. For instance, it is more efficient to use short wavelength for phase estimation, or a focused beam for position sensors. Optimal measurement and estimators are necessary to experimentally reach  the Cram\'er Rao bound.

When using light for optimal measurement, and before considering any purely quantum effects, one has to ascertain saturation of the QCRB with classical sources. 
This key point should become a strategy when optimising or proposing new imaging systems. It is so important that a research community has emerged around this concept alone. For example,  saturation of the QCRB is particularly relevant in the context of  parameters encoded in the modal structure of a light beam~\cite{fabre2020modes,gessner2023estimation}. These parameters{, typically considered in imaging devices,} do not change the quantum state but the underlying classical mode basis on which it is built. {They encompass} spatial intensity distribution, frequency spectrum, polarization and even phase of the light field. Measurement sensitivity {in these situations} directly depends on the {chosen} mode basis, and choosing {it} properly can induce a dramatic scaling change~\cite{gessner2020superresolution}.

\paragraph{Separating sources.} {Let's delve into a} simple case{:} transverse localisation of point-like incoherent optical sources. The transverse displacement of such sources corresponds to a displacement of the mode basis. A measurement device sensitive to the spatial mode proportional to the transverse derivative of the mean field is optimal, and can reach the QCRB~\cite{fabre2020modes,pinel2012ultimate}. Only then, using squeezing in this derivative mode, one can go beyond the classical limit~\cite{treps2003quantum}. This {concept extends} to the separation measurement of two incoherent sources, where modal decomposition plus intensity measurement reach{es} the QCRB~\cite{tsang2016quantum}{, as recently demonstrated experimentally~\cite{boucher2020spatial,santamaria2023spatial,rouviere2023ultra}}. {In the experiment of Ref. ~\cite{rouviere2023ultra}, illustrated in Figure~\ref{Figure2}, researchers used spatial-mode demultiplexing (SPADE)} {to enhance the} sensitivity five orders of magnitude beyond the Rayleigh limit,{practically reaching a sensitivity of 20 nm with an accuracy of 1 µm for a 1 mm beam size for bright sources.} {Even for faint sources, their approach {exhibited precision of 20 µm for a 1mm beam size with around 200 measured photons in the chosen mode, surpassing the} performance {of} ideal direct imaging methods (infinite resolution, no noise, equivalent losses). This experiment is a practical demonstration of passive imaging that significantly exceeds the Rayleigh limit using {SPADE}, {with} a straightforward setup adaptable to standard practices.} 

\paragraph{{Challenges and applications.}} Recent work has extended results to the time domain \cite{Silberhorn21} and, for spatial imaging, to the estimation of the 2D source position rather than just 1D separation \cite{Tan:23}. However, despite these extensions and therefore the first attempts to extend SPADE and related techniques, {localizing} multiple (>2) sources and/or {conducting} 2D imaging of spatially extended or complex shapes remains {challenging}. Although there {is not} a fundamental reason {prohibiting} QCRB-limited 2D imaging, the complexity of estimating the correct mode basis and subsequent{tly} {estimating} the full range of parameters, even for several point-like sources, is prohibitively difficult. Opportunities may therefore lie at the boundary of `hard imaging problems' and constrained-parameter estimation that still allow forms of SPADE superresolution.

\section{Imaging with {bright} squeezed light}
\label{part1}

{In most optical imaging approaches}, information is encoded in the intensity or phase of an optical field. Phase contrast and absorption microscopy are obvious examples, but this is also true for other imaging systems. For instance, fluorescence microscopy {encodes information} {in fluorescence intensity}, Raman microscopy {in the} intensity of the Stokes-scattered field, and polarisation microscopy {in the} relative phase shifts of polarisations.

\paragraph{{Beating shot noise.}} Quantum mechanics imposes limits {on simultaneously knowing} the intensity and phase of an optical field {due to} the Heisenberg uncertainty principle, {introducing} a fundamental source of noise in optical imaging. In the absence of quantum correlations between photons, this noise is {called} {\it shot noise} and {arises} from the random arrival of photons at the {detector}. Bright squeezed-light based imaging systems use nonlinear processes {like} optical parametric amplification~\cite{bowen2002recovery} and four wave mixing~\cite{mccormick2007strong} to manipulate {this} noise, {introducing} quantum correlations. While {respecting} the Heisenberg uncertainty principle, squeezing {reduces} noise {in} intensity at the expense of increased {phase} noise, or vice versa. {This reduction in noise  on the information-carrying variable enhances the signal-to-noise ratio (SNR), constituting a form of quantum-enhanced measurement.}

{Interest in} quantum-enhanced measurements using bright squeezed light {surged in} the 1980's, motivated especially by possible applications in gravitational wave detection, which requires {detecting} exceptionally weak signals~\cite{harry2010advanced}. This application has now been realised, with squeezed light routinely {improving} the noise floor of science runs in the LIGO and GEO600 gravitational wave observatories~\cite{dwyer2022squeezing}. While other sources of quantum light {offer} enhanced SNR, bright squeezed light was chosen for two primary reasons. First, 
squeezing is compatible with bright optical fields, thus overcoming the common `low-light' drawback in quantum sensing~\cite{taylor2016quantum}. Second, high levels of quantum enhancement are possible using standard photoreceivers, albeit with high-efficiency and low noise. Noise suppression of as much as 15~dB has been reported~\cite{vahlbruch2016detection}.

Like gravitational wave observatories, many imaging systems are used to detect faint signals against a large noise background. For instance, in microscopy, the smaller a structure the weaker its interaction with light and the harder it is to detect. A long running trend has been to use bright illumination sources to increase {signals strength}~\cite{taylor2016quantum}. However, the specimen {'s nature} ultimately constrains {the brightness of} illumination {it can withstand without degrading.}  This limits {imposes a} maximum {achievable} SNR without quantum correlations, especially strong for biological imaging where light can alter biological structures, {generate} toxic free radicals, affect function, and ultimately kill the specimen~\cite{taylor2016quantum}. 

Bright squeezed light is ideally suited to overcome the photo-damage limit to SNR in 
imaging {due to} its compatibility with bright optical fields and {its easy integration with} standard imaging systems and detectors~\cite{taylor2016quantum,lawrie2019quantum}. It has been applied {in various} imaging and imaging-related experiments, including optical tweezers~\cite{taylor2013biological, taylor2014subdiffraction}, atomic-force microscopy~\cite{treps2003quantum, pooser2020truncated}, plasmonic sensing~\cite{dowran2018quantum, pooser2016plasmonic}, Raman microscopy~\cite{casacio2021quantum,de2020quantum,xu2022stimulated,xu2022quantum,xu2023dual}, and Brillouin microscopy~\cite{li2022quantum}.

\paragraph{{Biological imaging with {bright} squeezed light.}} The first quantum-enhanced biological measurements using {bright} squeezed light were reported in Refs.~\cite{taylor2013biological, taylor2014subdiffraction}. Bright squeezed light reduce{d} optical noise associated with tracking {particle positions} in optical tweezer{s}, {increasing the} information extracted about {their} environment. This was applied to track lipid particles within a yeast cell {to} measure the local viscosity with quantum-enhanced sensitivity. As the particle explored the cell, it was possible to extract one-dimensional maps of {cellular} structures with {about} 10~nm {resolution,} enhanced by 14\% compared to measurements without squeezing. 

The first quantum-enhanced two-dimensional images generated using bright squeezed light were reported in Ref.~\cite{casacio2021quantum}. Squeezing was used to reduce the noise floor of a stimulated Raman microscope. {This} microscope uses a pair of bright different-frequency illumination fields~\cite{cheng2015vibrational,camp2015chemically} {, with} the frequency difference match{ing} the vibration frequency of a particular molecular bond, such as the C-H stretch bond. The presence of molecules {containing} that bond introduces scattering from one field into the other{, enabling} unlabelled chemically specific imaging. Stimulated Raman microscopy is a powerful technique {increasingly} used in biological imaging~\cite{cheng2015vibrational,camp2015chemically}. However, state-of-the-art microscopes already operate at the shot noise limit and photo-damage constrains the optical intensities~\cite{fu2006characterization}. {Thus, improving}  SNR requires new approaches, motivating the application of bright squeezed light in Ref.~\cite{casacio2021quantum}.

In Ref.~\cite{casacio2021quantum}, the authors imaged {lipid concentrations} in yeast cells, achieving a {35\%} improvement in SNR (see Figure~\ref{Figure2}). Importantly, {they} showed that {even a small increase in illumination intensity caused} irreversible {cell} damage. {Thus}, the experiment demonstrated a form of absolute quantum advantage{:} the {achieved} SNR {was unattainable} without quantum correlations. {However, }{this advantage was limited} to the {specific} apparatus and optical fields {used}. {Demonstrating apparatus-independent} quantum advantage {remains a significant challenge}.

\paragraph{{Challenges and applications.}} Bright squeezed light is {increasingly applied in imaging and} biological microscopy, {improving} SNR{, } resolution, and observing structures obscured {by} regular light sources. Major challenges remain to see broad impact from squeezing, {such as enhancing} noise suppression, demonstrating applications in {various} microscopes and imaging systems, and developing easy-to-use{, } robust methods {for} use outside of physics laboratories.

\section{Imaging with {two-photon states}}
\label{part2}

Previous sections have dealt with enhancements {using} squeezed states of light occupying a single spatial mode of the electromagnetic field.
However, squeezing can also occur between multiple modes.
Two-mode squeezed vacuum states, for example, are {easily} produce{d} through spontaneous parametric down conversion (SPDC)~\cite{burnham1970observation}. In the low-gain regime, these states can be approximated by entangled two-photon states. Such states provide an additional resource{, enabling the utilization} of correlated-photon imaging techniques. For instance, they can be used to achieve sub-shot-noise imaging, as demonstrated with low-absorptive~\cite{brida2010experimental,samantaray2017realization} and phase objects~\cite{ortolano2023quantum}. In such schemes, illustrated in Figure~B1.b, intensity fluctuations are {identical} at specific transverse position{s} on each side of the camera. {Consequently, } measurement noise in one {beam} {is} reduced by subtracting the signal intensity image from the idler intensity image.

However, photon-pair have applications beyond noise reduction. As illustrated in Figure~\ref{Figure4}a, recent advances in utilizing single-photon sensitive cameras to capture their quantum properties~\cite{moreau2019imaging,madonini2021single,vidyapin2023characterisation} have {spurred} innovative quantum imaging approaches. {These} include protocols addressing specific classical limitations, such as spatial resolution or phase sensitivity, enhancing technical aspects {like} background light reduction, and {introducing} novel 
imaging modalities with no classical counterparts.

\paragraph{Enhancing spatial resolution.} \label{resopair}
{Imaging} beyond the diffraction limit, especially in microscopy, reveals finer, previously unseen structures compared to conventional systems. Section~\ref{part0} {highlighted} how {optimizing} the detection mode basis can enhance imaging beyond the Rayleigh limit. {Additionally, }one can also optimise the input state. Multi-photon Fock states with {enable} correlation-based detection approaches{,} where simultaneous detection of $N$ photons enhances the transverse spatial resolution by a factor of $1/\sqrt{N}$, reaching the standard quantum limit (SQL)~\cite{giovannetti2009sub}. {Recent} non-fluorescent full-field imaging systems have {approached} this limit {with} photon pairs i.e. $N=2$~\cite{reichert2018massively,toninelli2019resolution}. This was achieved with Electron Multiplied Charge Coupled Devices (EMCCDs) {to detect} photon coincidences between adjacent pixels~\cite{reichert2018massively} or by using a centroid approach~\cite{toninelli2019resolution}. The resulting image is formed by convolving the object with a resolution-enhanced point spread function (PSF), {$h^2$}, where $h$ is the classical, coherent PSF. {This} enhanced-resolution approach was {recently} implemented in a bright-field transmission microscope~\cite{he2023quantum}, enabling the observation of cells with improved resolution and contrast at very low flux (Fig.~\ref{Figure4}.d). 

Harnessing the quantum coherence of an $N$-photon entangled state {allows} reaching the Heisenberg limit, providing a $1/N$ resolution enhancement~\cite{giovannetti2009sub}. This can be achieved by detecting photons in a diffraction plane, forming the foundation of quantum lithography~\cite{boto2000quantum, d2001two}. In practice, however, this approach is hampered by the {inefficiency of} multi-photon-absorbing materials. When imaging the object, a recent scheme reached this limit using photon pairs~\cite{unternahrer2018super}. {Here}, the object is placed in the pump beam's path {before} the crystal. The resulting coincidence image, captured with a single photon avalanche diode (SPAD) camera, results from convolving the object with an effective PSF, $[h* h](2 \textbf{r})$. This doubles the resolution based on the photon-pair wavelength, yet the object still requires illumination from the pump at half the photon-pair wavelength.

\paragraph{Improving phase sensitivity.} Phase imaging provides an alternative way to exploit the coherence of $N$-photon states, as seen in quantum metrology using $NOON$ states~\cite{giovannetti2011advances}. The de Broglie wavelength scales as $\lambda/N$, {providing} a $\sqrt{N}$-fold improvement in phase measurement sensitivity {over} classical interferometry. Initially {used} in scanning imaging systems~\cite{ono2013entanglement,israel2014supersensitive}, this concept was recently extended to full-field imaging with photon pairs{, }implemented with SPAD~\cite{camphausen2021quantum} and EMCCD~\cite{defienne2022pixel} cameras. In a related context, recent techniques exploit the Hong-Ou-Mandel effect across multiple pixels to enhance sensitivity when reconstructing transparent sample depth-profiles~\cite{devaux2020imaging,ndagano2022quantum}. As a photon passes through the sample, it experiences a group velocity delay, shifting the HOM dip and {varying} photon coincidence rates. This enables precise measurement of sample optical thickness at the nanometer scale~\cite{lyons_nm}.

{Note} that the sensitivity and resolution {improvements} mentioned in previous paragraphs are relevant only when comparing imaging systems with equal noise level{s}. Proper definitions must {consider} both optical system parameters (e.g. optical resolution, contrast) and measurement noise~\cite{tsang2016quantum}. Increasing photon flux is {indeed often} the best strategy for improving resolution and sensitivity, an aspect in which current $N$-photon sources are orders of magnitude inferior to classical sources. 

\paragraph{Reducing background noise.}
\label{noisepair}
Measuring photon coincidences {often poses} a technical challenge, {yet} it can effectively reduce background noise and stray light. {Known} as quantum illumination~\cite{lloyd2008enhanced}, this approach hinges on discerning the genuine coincidences between spatially entangled photons to distinguish {them} from accidental coincidences {stemming} from uncorrelated classical sources. Originally developed for target detection~\cite{lopaeva2013experimental}, this concept has recently {been adapted for} full-field imaging setups employing entangled photon pairs that can differentiate or `distil' quantum from classical images~\cite{defienne2019quantum,gregory2020imaging}. As depicted in Figure~\ref{Figure4}. {f}, this separation is achieved through covariance measurements~\cite{defienne2019quantum} or photon-pair identification~\cite{gregory2020imaging} with EMCCDs. This approach has garnered significant interest in recent years for Light Detection and Ranging (LiDAR) applications. Recent studies have explored scenarios {where} a scene is illuminated {using either} entangled photons~\cite{zhao2022light} or heralded {single-photons}~\cite{england2019quantum}. By measuring correlations in the scattered light, this protects the received signal from ambient light or jamming, a concept that also inspired classical implementations~\cite{liu2023compact}. Such stray light robustness was also {recently} used in a quantum scanning {microscope}~\cite{zhang2023quantum}.

\paragraph{Developing new modalities.}
\label{newpair}
Quantum light also enables new imaging modalities. Like the pioneering works on
% \cite{white1998interaction}
Ghost imaging~\cite{pittman1995optical} (Box 1), these schemes are primarily driven by fundamental exploration rather than a specific need to surpass classical boundaries. Recent examples include quantum holography approaches {utilizing} photon pairs~\cite{devaux2019quantum,defienne2021polarization}. One {such} schemes, shown in Figure~\ref{Figure4}. {i}, leverages polarization entanglement for non-local holographic reconstruction of phase objects, offering advantages including enhanced resolution, noise robustness and resistance to random phase distortions. This relies on {maintaining} correlations across orthogonal polarization bases, underscoring the essential role of entanglement. Other examples encompass photon-pair-based methods for hyper-spectral imaging~\cite{zhang2023snapshot}, multimodal microscopy~\cite{hodgson2023reconfigurable}, adaptive optics~\cite{cameron2023quantum}, non-degenerate~\cite{aspden2015photon} and entanglement-swapped-based Ghost imaging~\cite{bornman2019ghost}, as well as schemes based on induced coherence~\cite{lemos2014quantum}, detailed in section~\ref{part3}. 

\paragraph{Challenges and applications.}
Despite recent progress, multi-photon Fock-state-based quantum imaging approaches have yet to outperform classical imaging techniques in practical applications. This limitation is primarily rooted in the sources employed. SPDC faces efficiency challenges when producing states with $N > 2$. Even for $N=2$, the fluxes are typically in the nanowatt range. Alternative technologies, such as semiconductor quantum dots~\cite{somaschi2016near}, are under development and show promise. {Additionally}, significant progress is needed on the detector front to shorten acquisition time. Event-based cameras, exemplified by recent SPAD~\cite{madonini2021single} and intensified camera~\cite{vidyapin2023characterisation} models, {are among} the most encouraging technologies. As we await these developments, current quantum imaging systems mainly {find potential} in niche applications. {Promising} avenues include biological imaging in photo-sensitive samples~\cite{he2023quantum,zhang2023quantum} (Fig.~\ref{Figure4}. {d}) and in wavelength ranges that are difficult to detect~\cite{lemos2014quantum,aspden2015photon}.

\section{Imaging with undetected photons}
\label{part3}

Among developments involving SPDC sources, `imaging with undetected photons (IUP)’, {also known} as `coherence induced imaging', shows strong {application} potential {by decoupling} the sensing and detection wavelengths. Illustrated in Figure~B1.c, the measurement scheme uses a nonlinear interferometer comprising two nonlinear crystals (S1 and S2) coherently pumped in the low-gain regime, {where} stimulated emission is negligible {compared to} spontaneous emission. Individually, the resulting signal and idler fields thus generated would have {random} or undefined  phases{, preventing}  first-order interference between the signal (or idler) fields. However, if one seeds the idler mode in both crystals with the same light field, {it induces} stimulated emission of {phase-locked} signal fields{, allowing} first-order coherence {and interference} between the signal fields of both crystals~\cite{wang1991observation}. This scheme was used for imaging in {Ref.}~\cite{Cardoso2017}. The ‘seeded’ processes{, known as} optical parametric amplifiers (OPA), have a classical explanation. Surprisingly, first-order coherence can still be fully preserved even without seeding and in the low gain regime
 
For this, the idler spatial mode from the first crystal must perfectly overlaps with that of the idler photon emitted in the second crystal, and certain path length criteria must be satisfied~\cite{lemos2022quantum}. Even if the emission rates are now so low that the probability of both crystals emitting a pair at the same time and stimulated emission is negligible, first-order coherence is fully preserved. {Classical models fail} to explain this phenomenon{, requiring a} quantum description - this is the case of IUP~\cite{lahiri2019nonclassicality}. Discovered by Ou, Wang, Zou and Mandel~\cite{wang1991}, {they} provided a quantum mechanical explanation that uses the path indistinguishability of the photons emitted from the two crystals. They showed that {placing} a partially absorbing or reflecting object in the idler field between S1 and S2 reduces the interference fringe visibility {of combined} signal fields from S1 and S2, even {without detecting} the idler field. No coincidence detection is required. This concept was later generalized to multi-mode phase and absorption imaging by Lemos et al.~\cite{lemos2014quantum}. {This} imaging method can also operate in the high-gain regime~\cite{shapiro2015}, i.e. induced coherence with stimulated emission, but without external seeding. The advantages of {each regime must} be considered case by case~\cite{kolobov2017, lemos2022quantum}.

A key feature of IUP is that the idler photon from S1 {interacting} with the object can be at any desired wavelength {and} must never be detected{.} This removes all constraints on the detector technology{, as long as the} signal photons are easily {detectable,} typically in the visible or near-infrared region. The idler photon can then be used, for example, for imaging at mid-infrared wavelengths, {beyond} the sensitivity of Si-based camera{s,} for which cost-effective, multi-pixel sensors with low-light sensitivity and reasonable SNR and/or spatial resolution are hard or impossible to purchase. Most recent implementations of IUP {use} a `folded' setup {instead of two independent crystals, as illustrated in Figure~\ref{Figure5}a.} 

\paragraph{Spatial resolution and field-of-view.}
Besides choosing advantageous wavelength combinations for IUP, resolution and field-of-view (FOV) are crucial parameters {in real-world applications}. As with other imaging schemes using photon pairs, IUP uses transverse spatial correlations between signal and idler photons. Formulas for the ideal spatial resolution{,} derived from the quantum description of spatially entangled photon pairs~\cite{viswanathan2021position,fuenzalida2020resolution,kviatkovsky2020microscopy,basset2023experimental}{, }are shown in Table~1 together with {some} recent experimental results. The derivations assume that the optical components do not impose stringent diffraction limits or partial obstructions to the beams. {Generally,} spatial resolution is diffraction-limited to the longer wavelength of the photon pair~\cite{fuenzalida2020resolution}. This can be side-stepped by using near-field correlations {with} ultra-thin crystals~\cite{santos2022subdiffraction}, also strongly reducing the rate of generated pairs.
{In} IUP{,} distinct imaging conditions {include} the far-field (FF) {and the image-plane (IP) configurations.} {In the FF configuration,} {where} the imaged object and camera are placed at the Fourier-plane of both crystals{, the pump waist $w_{0p}$ determines the resolution $res$, while the crystal length $L$ determines the field of view (FOV).} In {the IP} configuration{, where} the sample and camera are in the image-plane of both crystals{, the pump waist affects the FOV and the crystal length affects the resolution}. {Any} specific resolution{s} (within the diffraction limit of the longer wavelength) can be obtained by appropriate magnification optics. However, the ratio $N={FOV}/{res}$, {indicating} the number of 1D-resolvable spatial elements within the FOV, is independent of magnification. $N$ is maximized for short crystals and large pump-waist {but decreases with} {a} larger ratio {$\lambda_i/\lambda_s$, where $\lambda_i$ and $\lambda_s$ are the signal and idler wavelengths, respectively~\cite{kviatkovsky2022mid}.} {Typically,} $N$ {is large} for idler wavelengths in the Mid-IR {, while it becomes} impractically small for THz idler photons~\cite{kutas2020terahertz}.

\begin{table}
    \centering
    \begin{tabular}{rcccccccl}
    \rowcolor{darkgreen}
     \multicolumn{9}{l}{ \fontsize{11}{16} \textbf{Table 1 | Resolution and field-of-view constraints for imaging with undetected photons.                   } }  \\
    %\hline 
         \rowcolor{lightgreen} &  \multicolumn{4}{c}{\textbf{FF-configuration}} & \multicolumn{4}{c}{\textbf{IP-configuration}} \\
    \hline \rowcolor{lightgreen} 
         \textbf{Resolution (FWHM)} &  \multicolumn{4}{c}{$res_{FF}=f{\sqrt{2Ln(2)/\pi^2}}\:\lambda_i/w_{0p}/M$} & \multicolumn{4}{c}{$res_{IP}=0.44\sqrt{L}\sqrt{(n_s\lambda_i+n_i\lambda_s)/(n_s n_i )}/M$} \\
    \rowcolor{lightgreen} 
         \textbf{$FOV$ (FWHM)} &  \multicolumn{4}{c}{$FOV_{FF}=1.88\lambda_i\frac{f}{\sqrt{L}}\sqrt{n_s n_i/(n_s\lambda_i+n_i\lambda_s)}/M$} & \multicolumn{4}{c}{$FOV_{IP}=\sqrt{2ln2}\: w_{0p}/M$} \\
    \rowcolor{lightgreen} 
         \textbf{$N$ (1D number of modes)} &  \multicolumn{4}{c}{$N_{FF}=5.0\frac{w_{0p}}{\sqrt{L}}\sqrt{n_s n_i/(n_s\lambda_i+n_i\lambda_s)}$} & \multicolumn{4}{c}{$N_{IP}=2.7\frac{w_{0p}}{\sqrt{L}}\sqrt{n_s n_i/(n_s\lambda_i+n_i\lambda_s)}$} \\
        \rowcolor{lightgreen} &&&&&&&& \\
         \rowcolor{lightgreen}  \textbf{Experiments} &  \textbf{$FF/IP$} &  \textbf{$\lambda_i$ ($\mu m$)} &  \textbf{$\lambda_s$ ($nm$)} &  \textbf{$L$} (mm) &  \textbf{$w_{0p}$ ($\mu m$)} &  \textbf{$N$} (1D) &  \textbf{$FOV$} (mm) & \textbf{$res$ ($\mu m$)} \\
    \hline \rowcolor{lightgreen} 
         2014, Lemos et al. \cite{lemos2014quantum} &  FF &  1.55 &  810 &  2 &  200 & 21  & 5.0  & 235 \\
    \rowcolor{lightgreen} 
         2020, Paterova at al. \cite{paterova2020hyperspectral} &  FF &  2.8-3.4 &  630-650 &  10 &  400 &  32 &  0.55 & 17 \\
    \rowcolor{lightgreen} 
         2020, Paterova et al. \cite{paterova2020quantum} &  FF & 1.55 & 810 & 10 & 500 & 62 & 2.4 / 0.5 & 39 / 7.8 \\
    \rowcolor{lightgreen} 
         2020, Kviatkovsky et al. \cite{kviatkovsky2020microscopy} &  FF &  3.7 &  800 &  2 &  430 &  23 &  0.82 & 35 \\
    \rowcolor{lightgreen} 
         2020, Fuenzalida et al. \cite{fuenzalida2020resolution} & FF & 1.55 & 810 & 2 & 180 & 10 & 1.8 & 180\\
    \rowcolor{lightgreen} 
         2021, Basset et al. \cite{gilaberte2021video} & FF & 730 & 910 & 2 & 415 & 42 & 2.515 & 60 \\
    \rowcolor{lightgreen} 
         2022, Kviatkovsky et al. \cite{kviatkovsky2022mid} & IP & 3.7 & 800 & 2 & 430 & 18 &  0.16 & 9\\
    \rowcolor{lightgreen} 
         2023, Pearce et al. \cite{pearce2023practical} & FF & 1.45 & 840 & 10 & 500 & 20 &  4.0 & 200\\
    \rowcolor{lightgreen} 
         2023, Basset et al. \cite{basset2023experimental} & IP & 910 & 730 & 2, 5 & 305 & 20, 12 & 0.32 & 16, 26 \\
         \hline
    \rowcolor{lightgreen} \multicolumn{9}{l}{ \fontsize{8}{8} \textit{FF: far-field configuration. Object and camera positioned at the Fourier-plane of the first and second crystal, respectively.}} \\
    \rowcolor{lightgreen} \multicolumn{9}{l}{ \fontsize{8}{8} \textit{IP: Image-plane configuration. Object and camera are in the image-plane of the first and second crystal, respectively.}}\\
    \rowcolor{lightgreen} \multicolumn{9}{l}{ \fontsize{8}{8} \textit{$\lambda_s,\lambda_i$: signal and idler wavelength, respectively; $n_s,n_i$: crystal refraction indices for $\lambda_s$ and $\lambda_i$, respectively.} }\\
    \rowcolor{lightgreen} \multicolumn{9}{l}{ \fontsize{8}{8} \textit{$w_{0p}$: pump waist; $L$: crystal length; $M$: system magnification.} }
    \end{tabular}
    \label{table}
\end{table}

\paragraph{Extension {to other imaging modalities}.}
{IUP was extended to} quantitative phase imaging by {phase-shifting} the signal or idler arm and extracting {the} relative phase value at each pixel~\cite{kviatkovsky2020microscopy,fuenzalida2020resolution,topfer2022quantum}.
3D-imaging can {also} be {achieved} {with} IUP{-based} optical coherence tomography (OCT), demonstrated first in the time-domain (TD-OCT)~\cite{valles2018optical,paterova2018tunable} and later in the frequency-domain (FD-OCT)~\cite{vanselow2020frequency}. Spectral noise{, }the dominating limitation for the SNR of FD-OCT, is fundamentally suppressed for SPDC{, allowing} {IUP-based} FD-OCT to reach shot-noise-limited sensitivity \cite{vanselow2020frequency}. High-gain regime  FD-OCT with undetected photons was {also} demonstrated\cite{machado2020optical}, exhibiting loss-resilient sensitivity.
Another example is spectral imaging which {is highly relevant for application}, especially in the mid-IR. Here, the abundant spectral and transverse spatial entanglement in SPDC can be leveraged. {For example,} the absorption spectrum of $CO_2$ at 4.3$\mu m$ was obtained through the detection of visible light~\cite{kalashnikov2016infrared}. This could be extended to spectral imaging by tuning the {SPDC} wavelength~\cite{paterova2020hyperspectral} or by using variable spectral filtering~\cite{kviatkovsky2020microscopy}. Other approaches include fast, grating-spectrometer based spectroscopy~\cite{kaufmann2022mid}, which can be combined with point scanning{, and} wide-field Fourier-transform imaging~\cite{placke2023fourier}.

\paragraph{{Challenges and applications.}}
While IUP has already been demonstrated for {various} modalities from microscopy to 3D-tomography and spectral imaging\cite{lemos2022quantum}, it {has yet to become a} commercial technology. A first target commercialisation area could be for mid-IR imaging, where {IUP offers} advantages {in} foot-print, complexity, cost or measurement-speed. Applications could {include} Si-chip inspection, tissue imaging, micro-plastic identification and non-destructive testing of ceramics. {To make IUP competitive,} a general challenge is to significantly increase the source brightness. {Strategies include}  leveraging high peak power pulsed lasers~\cite{hashimoto2023broadband} {to} push the SPDC process into the high-gain regime{, }employing cavity-enhancement for the pump laser~\cite{lindner2023high} or significantly increasing the  length of non-linear crystals{, } both of which can boost brightness by 1-2 orders of magnitude. Seeding with a stronger light field in the signal (or idler) mode, such as in Ref.~\cite{Cardoso2017}, {is another} fruitful direction to explore further.  A general  advantage of IUP is that it can be made very compact~\cite{pearce2023practical} and could therefore enable mobile devices for commercially interesting application{s}{, } and even implementations on-chip can be envisioned.

\section{Imaging with single photon {states}}
\label{part4}

One factor hindering broader implementation of quantum optical methods in microscopy is the apparent requirement for a source of non-classical light. These sources are often considered beyond the reach of a common bio-imaging laboratory. {However}, several schemes have been proposed for superresolving statistically independent emitters only through intensity correlation measurements, obviating the need for a special light source~\cite{thiel2007quantum,oppel2012}. This raises the question: can a practical quantum advantage be achieved {with} current microscopes? The perhaps counter-intuitive answer to this question is a clear yes. This is because most fluorophores used in bio-imaging are ‘quantum emitters’ incapable of simultaneously emitting a pair of photons. A dye molecule for example, relaxes to its electronic ground state after emitting a photon and needs time to re-excite before emit{ting} {another}. This property{,} known as ‘photon antibunching’{,} has no classical counterpart. Quantumness is {evident as} photons are more evenly spaced in time than classically allowed. A histogram of the number of emitted photons  per unit time {shows} variance less than the shot noise {predicted} by Poisson statistics. {Measuring this requires} at least a pair of single photon detectors in a Hanbury Brown and Twiss (HBT) setup, as shown in Figure~\ref{Figure5}.f. The light from the emitter is split using a beamsplitter to two (or more) single photon detectors, and the coincidence rate is measured (Fig.~\ref{Figure5}.g). To be practical, the {detectors' }time resolution should {match} the {emitter's }fluorescence lifetime, typically several nanoseconds.

\paragraph{Antibunching gives rise to a resolution enhancement.}
Photon antibunching can be used as a resource for enhancing the optical resolution of an imaging system. Antibunching is a quantum analog of stochastic classical intensity fluctuations {used in} superresolved imaging {methods} {like} Super-resolution Optical Fluctuation Imaging~\cite{dertinger2009fast}. Both classical and quantum fluctuations {deviate} from Poisson statistics of the number of photons emitted per unit time. Classical fluctuations, {like} fluorophore blinking, lead to super-Poissonian statistics whereas antibunching leads to sub-Poissonian statistics. {For} resolution, however, only the deviation from Poisson statistics matters since the higher moments of the distribution are not determined by the mean photon count. The deviation from Poisson statistics of the N$^\textrm{th}$ moment of the distribution, characterized by the correlation between detections from $N$ detectors, is associated with raising the PSF to the power of N {, resulting in approximately} $\sqrt{N}$ resolution enhancement. Yet, since in reciprocal space the spatial frequency support of the PSF is N times broader, the resolution can be further increased up to a limit determined by the SNR but not exceeding N times. Using pair correlations{, } the resolution increase is similar to that obtained {using} centroid estimation with SPDC photon pairs in transmission imaging \cite{unternahrer2018super,toninelli2019resolution}, described in section~\ref{resopair}. 

\paragraph{Direct use of antibunching for superresolution.}
The main difficulty in implementing antibunching-based superresolution imaging relates to the availability of suitable detectors. Estimating antibunching requires numerous binary frames (a photon detected or not at each pixel) to form an image. For wide-field imaging, a large format (of the order of 1 megapixel) single-photon sensitive sensor is necessary. The first demonstration of this method used an EMCCD detector operated in Geiger mode, limited to a readout rate of 1 kHz \cite{schwartz2013superresolution}. At such a low frame rate, the sample {must} be excited at a rate of one pulse per frame by pulses shorter than the radiative decay lifetime of the fluorescent marker (to avoid re-excitation) leading to {impractical} imaging times in any bio-imaging scenario. When much faster single photon imaging detectors with several tens of pixels became available (e.g. a fiber bundle camera~\cite{israel2017quantum} or a SPAD array~\cite{madonini2021single}), antibunching was used to augment the spatial resolution in confocal imaging~\cite{monticone2014beating} and in image scanning confocal microscopy~\cite{tenne2019super,lubin2019quantum}{. The latter} {achieved} up to threefold resolution increase beyond the diffraction limit by combining the resolution increase from image scanning microscopy and that from photon statistics. An example of this is shown in Figure~\ref{Figure5}.h, comparing a confocal image of labeled microtubules with a quantum-enhanced image. In this case, the improvement in lateral resolution is accompanied by an improvement in axial resolution and pixel dwell times could go as short as a few milliseconds~\cite{Sroda20}. Other superresolution modalities {like} Structured Illumination Microscopy~\cite{classen2017superresolution}, random illumination microscopy \cite{liu2022resolution} or Spatial Frequency-Modulated Imaging~\cite{field2016superresolved} can also similarly benefit from antibunching data provided appropriate detectors and data analysis.

\paragraph{Using photon statistics in localization based superresolution microscopy.} Fluorescent labels in bio-imaging{, } {behaving as} quantum emitters{, } can provide estimates on emitter numbers in a diffraction limited volume. The magnitude of the antibunching dip in the second-order photon correlation function $g^{2}(0)$ depends on the number of (assumed identical) emitters $N$ as:
$g^{2}(0)=1-1/N$, with extensions to higher orders \cite{weston2002measuring}.
Currently, localization based superresolution is limited to relatively sparse scenes since overlapping emitters introduce {reconstruction} artifacts. Considering localization microscopy as an optimization problem while using the estimated number of emitters as input to the localization algorithm{, these artifacts can be avoided.} {It enables} operation at higher emitter densities{, reducing} the number of acquired frames and {expediting} the overall image acquisition process~\cite{israel2017quantum}. 

\paragraph{Challenges and applications.}
The use of photon statistics has gained ground in many spectroscopy applications in the past decade, driven to a great degree by developments in detector technology. {Currently}, high-performance small detector arrays are sufficiently evolved to enable implementations in laser scanning microscopy (e.g. confocal, image scanning, stimulated emission depletion microscopy) or in applications requiring a very small field of view. {These detectors} also offer access to fluorescence lifetime information which is often useful. {However,} technology {for} widefield imaging {is still not fully developed}, {although} significant progress is {underway}, both in SPAD arrays \cite{wayne2022500} and in Superconducting Nanowire Single-Photon Detector (SNSPD) arrays~\cite{resta2023gigahertz}. Equally important is {the} develop{ment of} analysis algorithms {capable of efficiently} extract{ing} information from the (typically noisy) antibunching signal{s}, using both physical constraints {, }such as the intensity image \cite{rossman2019rapid} {,} and advanced data analysis schemes {like} deep learning \cite{kudyshev2023machine}. Combined, these will likely broaden the use of antibunching in imaging as a change of the detector and data analysis is typically the only required modification from the 'standard' microscope configuration, making it readily {applicable} in bio-imaging.

\label{fluo}

\section{Outlook}

Historically, optical imaging developed without {much consideration for} the quantum nature of light. This paradigm has shifted in the last 40 years, during which quantum imaging protocols evolved from scientific curiosities towards concrete applications. In their pursuit of enhancing imaging system performance, researchers have harnessed various non-classical states of light, coupled with advanced detection and manipulation methods to, for instance, surpass the resolution limit~\cite{tenne2019super}, capture images below the shot noise~\cite{treps2002surpassing}, operate in the presence of stray light~\cite{defienne2019quantum}, or work at virtually undetectable wavelengths~\cite{lemos2014quantum}. Under highly specific conditions and samples, recent quantum microscopy schemes have even unveil{ed} biological structures obscured by noise in classical approaches~\cite{casacio2021quantum}.

Still, it is important to underline that biologists are far from integrating quantum microscopes into their toolkit {due to several practical limitations, including} prohibitive acquisition times, very low illumination intensities and complex setup design. Nevertheless, the future holds great promise. Fueled by the second quantum revolution, recent technological advances {suggest} that quantum imaging systems will soon achieve the necessary performance levels for practical applications. Quantum light sources are becoming brighter, purer, involve more photons or higher squeezing levels~\cite{somaschi2016near}. {Meanwhile}, detectors are improving in sensitivity, speed, spatial and temporal resolution, featuring technologies like SPAD cameras~\cite{madonini2021single} or SNSPD arrays~\cite{resta2023gigahertz}.

As these technologies mature, one can envision an initial stage where quantum imaging systems complement classical approaches. This could {lead to} hybrid quantum-classical imaging, {with} conventional microscopes {incorporating} a 'quantum' mode to gain advantage under specific imaging conditions. Looking {ahead}, we anticipate a genuine transformation in optical imaging where quantum aspects will naturally be considered in the development of new tools. As global research funding organizations emphasize the advancement of practical quantum technologies, we {believe} that quantum imaging will be a significant contributor to this progress.

\section*{Acknowledgements}
W.P.B. acknowledges support from the Air Force Office of Scientific Research under award numbers FA9550-20-1-0391 and FA9550-22-1-0047, the Australian Research Council Centre of Excellence for Engineered Quantum Systems (EQUS, CE170100009), and the Australian Research Council Centre of Excellence in Quantum Biotechnology (QUBIC, CE230100021). H.D. acknowledges funding from the ERC Starting Grant (Grant No. SQIMIC-101039375). GBL acknowledges CAPES, CNPq, FAPERJ (JCNE, E-26/201.438/2021), from the John Templeton Foundation ( Grant No. 62424). SR acknowledges funding by the Federal Ministry of Education and Research (BMBF) under project numbers 13N16384, 13N15402 and 13N15944, as well as by the Einstein Foundation Berlin (EJF-2021-681). D.F. is supported by the Royal Academy of Engineering through the Chairs in Emerging Technologies programme and the UKRI Frontier Research scheme.

{\section*{{Competing interests}}
The authors declare no competing interests.

\section*{{Additional information}}
\textbf{Correspondence} should be addressed to H.D.}

\section*{{Author contributions}}

All the authors contributed to writing the article.

%\bibliography{Biblio}

%% bibliography

%% Figures

\newpage

\begin{figure}[!t]
\centering
\includegraphics[width=0.65\linewidth]{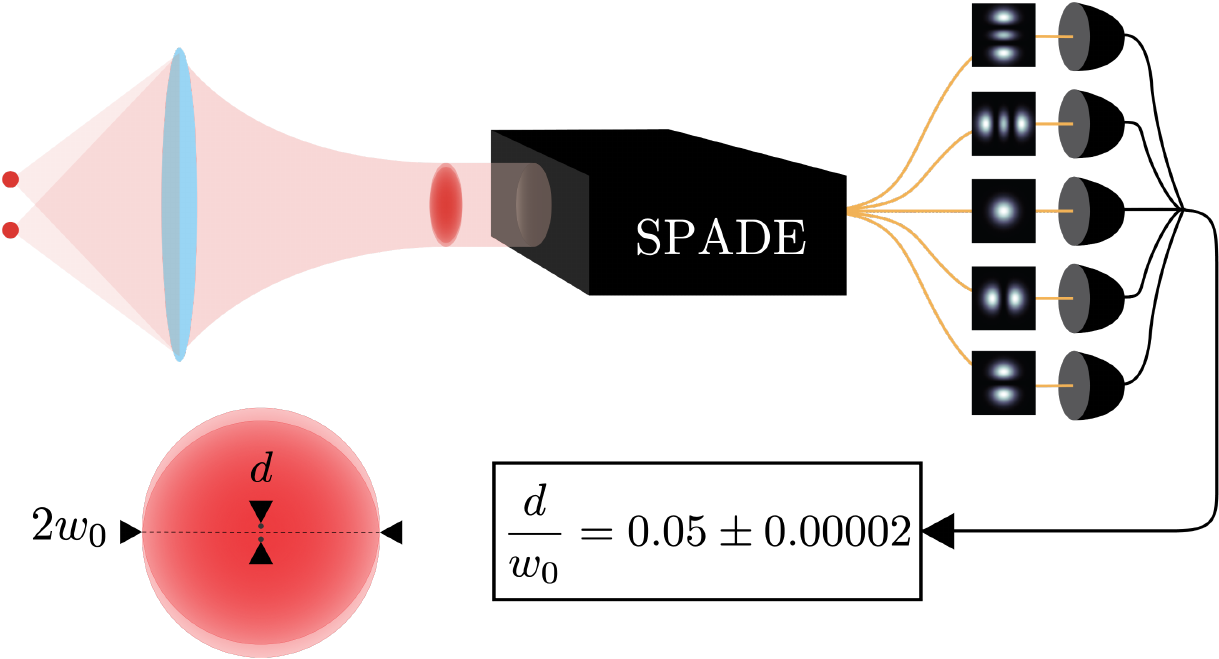}
\caption{{\textbf{Spatial-mode demultiplexing (SPADE) for optimal separation estimation.} SPADE  {decomposes} incoming light using the Hermite-Gaussian mode basis. Measuring the intensity {of} each mode is an optimal approach for transverse separation estimation. It provides orders of magnitude improvement in sensitivity for estimating closely spaced incoherent sources compared to direct diffraction-limited {imaging}. The numbers presented in this figure represent typical outcomes obtained using the {experiment} described in Ref.~\cite{rouviere2023ultra}, where $d$ denotes the transverse {beam} separation in the image plane, and $w_0$ represents the {beams} waist in the same plane.}} 
\label{Figure3}
\end{figure}

\begin{figure}[ht]
\centering
\includegraphics[width=0.6\linewidth]{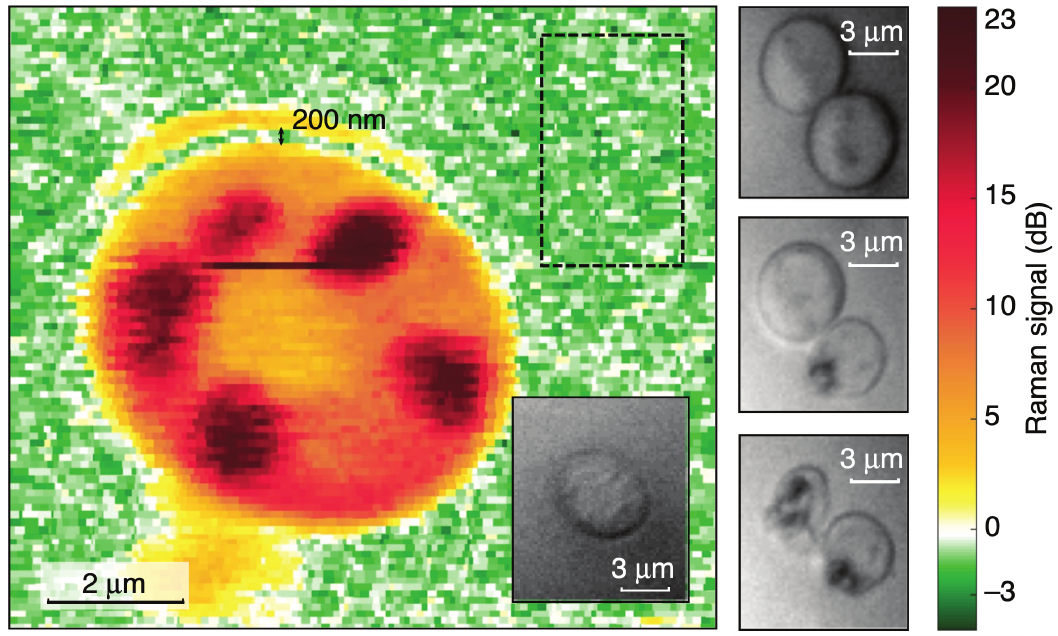}
\caption{\textbf{Biological imaging with {bright} squeezed light.}  A quantum-enhanced image of a live yeast cell using bright squeezed light in stimulated Raman microscopy~\cite{casacio2021quantum}. The frequency difference between the two illumination lasers was {set to} 2,850~cm$^{-1}$, a frequency {yielding} a strong Raman signal {from lipids}. A signal-to-noise {(SNR)} enhancement of 35\% was reported compared to the {SNR} achievable using shot noise limited light. On the right{, }a sequence of images {depicts} two cells illuminated with approximated twice the light intensity, showing visible photodamage after exposure for only a few seconds. }
\label{Figure2}
\end{figure}

\begin{figure}[ht]
\centering
\includegraphics[width=0.75\linewidth]{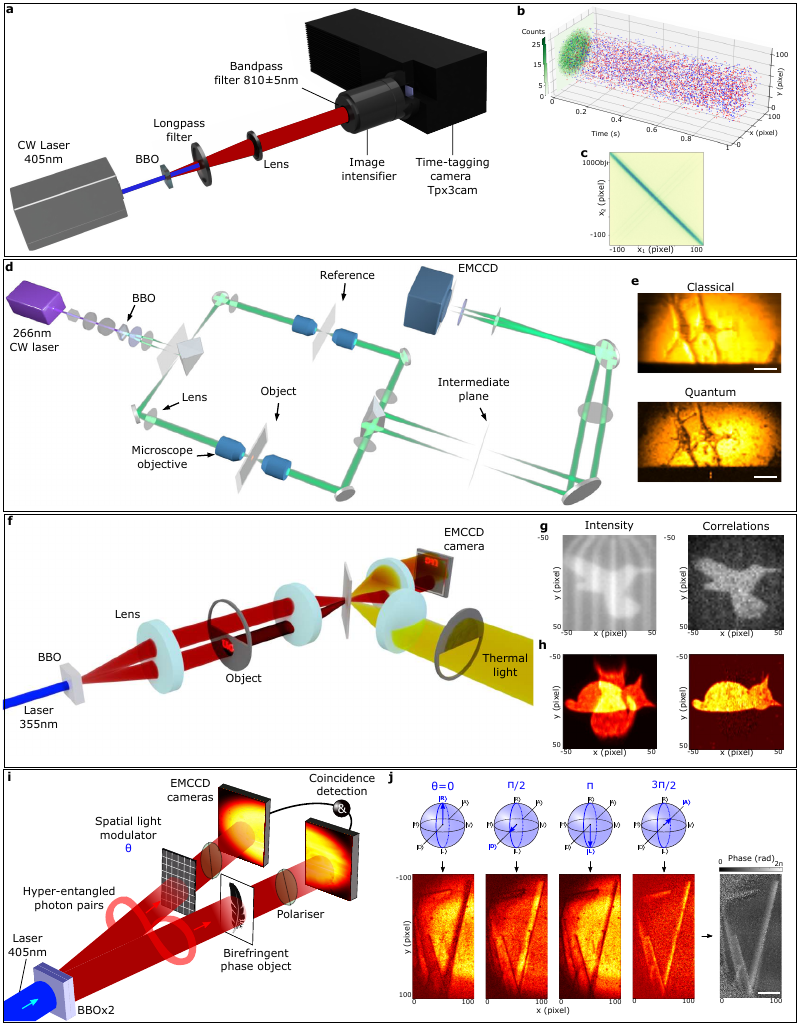}
\caption{\textbf{Imaging with {two-photon} states.} \textbf{a,} Imaging technique for detecting photons coincidences~\cite{vidyapin2023characterisation}. Photon pairs from SPDC strike an image amplifier, converting {each} photon into a flash of light {recorded} {by a} time-tagging camera. {\textbf{b, }}After cluster identification, the data contains all coincidence events (blue and red dots) detected over time per pixel {(}data from Ref.~\cite{courme2023quantifying}{)}.{\textbf{c, }}Properties such as strong spatial anti-correlations can thus be extracted. 
\textbf{d,} Optical microscope based on entangled photons and correlation measurement~\cite{he2023quantum}. \textbf{e,} Image{s} of HeLa cell acquired by classical and quantum microscopy. {Scale bar, 20$\mu m$.}
{{\textbf{f}},} Experimental realization of quantum illumination~\cite{gregory2020imaging}. {
\textbf{g}, The stray light (cage) superimposed on the object (bird) in the intensity image almost disappears when the image is reconstructed from the spatial correlations of the photons. } {\textbf{h, }} Similar {results are also obtained} by placing an object (sleeping cat) in a conjugate plane of the crystal~\cite{defienne2019quantum}. {{\textbf{i},}} Quantum holography enabled by entanglement~\cite{defienne2021polarization}. One photon from an entangled pair illuminates a birefringent phase object (bird feather), while the phase $\theta$ of its twin is controlled remotely using a spatial light modulator.\textbf{ j,} By varying $\theta$, the state {is measured in different polarization bases}, enabling reconstruction of the object's phase image. {Scale bar, 0.5mm.}}
\label{Figure4}
\end{figure}

\begin{figure}[ht]
\centering
\includegraphics[width=0.75 \linewidth]{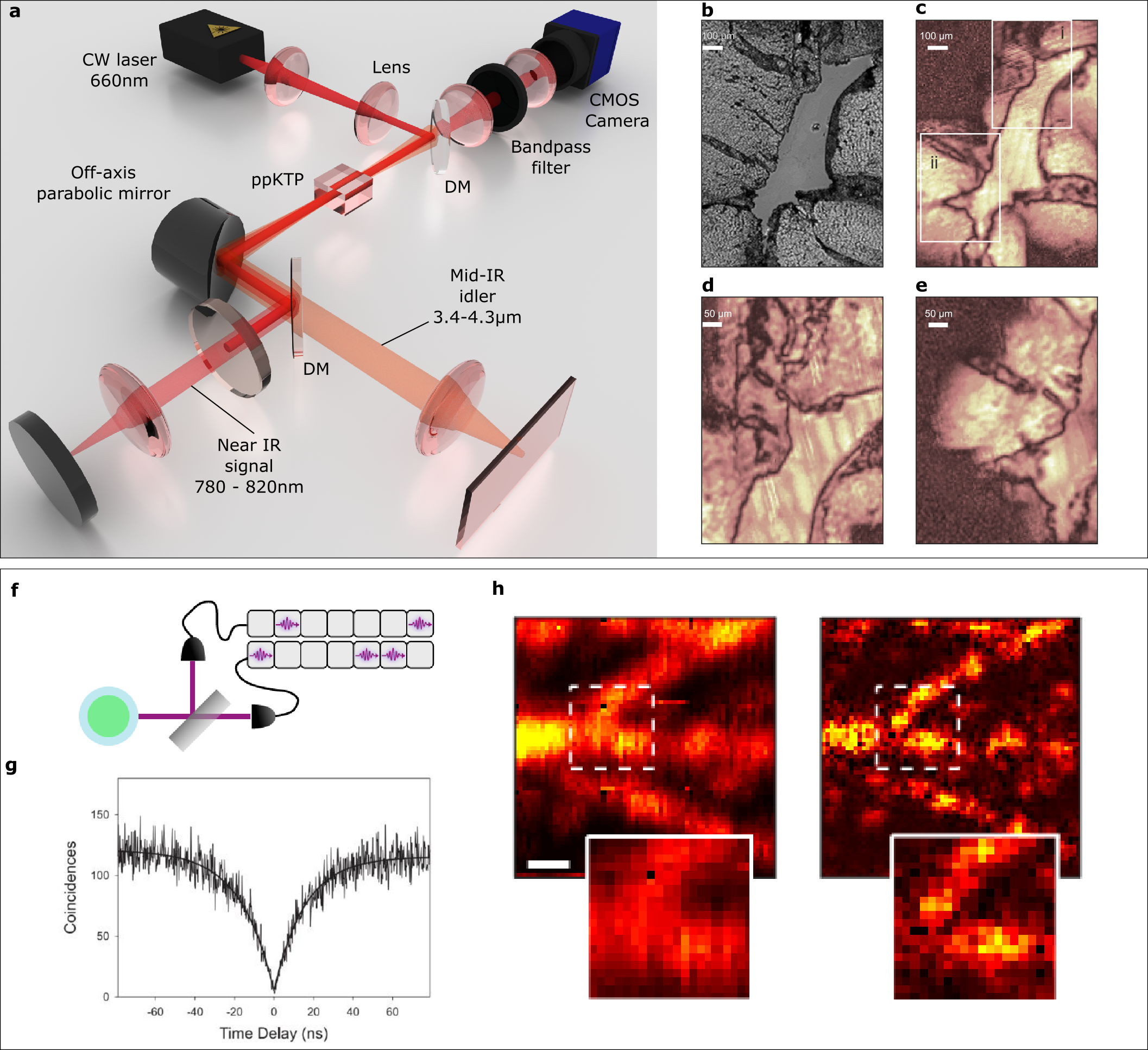}
\caption{{{\textbf{ Microscope{s} based on imaging with undetected photons (a-e)  {and single photon emitters (f-h)}}}}. {\textbf{a,} {Reflected light microscope based using a `folded' imaging with undetected photon configuration~\cite{kviatkovsky2022mid}. The sample is illuminated with mid-IR light while the image is formed onto the camera using near-IR.} \textbf{b,} Image of a mouse heart sample {acquired using conventional} bright-field microscopy with visible light. \textbf{c, d} and \textbf{e}, Mid-IR microscopy of the identical sample using undetected photons for absorption imaging. \textbf{f,} Detecting quantum emission from an individual nanoscale emitter accomplished using a Hanbury-Brown-Twiss (HBT) measurement~\cite{lounis2000photon}. \textbf{g},} The absence of coincidence detections, {analyzed} through the second-order correlation function $G^{(2)}(\tau)$, {indicated} photon antibunching {experimentally}.\textbf{ h,} Image scanning microscopy (ISM) and Quantum-ISM images obtained from a scan of microtubules in a fixed cell labeled with
fluorescent quantum dots~\cite{tenne2019super}. Following a confocal scan, $G^{(2)}(\tau)$ is measured between pairs of detectors for every scan step. The zero-delay dip at $G^{(2)}(\tau=0)$ serves as the contrast of Q-ISM. Scale bar is 0.5 $\mu$m.}
\label{Figure5}
\end{figure}

\newpage

\end{document}